\documentclass[sn-mathphys,pdflatex,iicol]{sn-jnl}

\usepackage{hyperref}
\usepackage{xcolor}
\usepackage{amsmath}
\usepackage{amsfonts}

\usepackage{mathtools}
\usepackage[noabbrev]{cleveref}

\DeclareMathOperator*{\argmin}{\text{arg min}}

\newcommand{\cd}{\cdot}
\newcommand{\cX}{\mathcal{X}}
\newcommand{\cD}{\mathcal{D}}

\newcommand{\rbr}[1]{\left(#1\right)}
\newcommand{\sbr}[1]{\left[#1\right]}
\newcommand{\ceq}{\coloneqq}

\jyear{2021}%

\theoremstyle{thmstyleone}%
\theoremstyle{thmstyletwo}%

\theoremstyle{thmstylethree}%

\begin{document}


\title[Article Title]{Deep Learning of Dynamical System Parameters from Return Maps as Images}


\author[1]{\fnm{Connor James} \sur{Stephens}}\email{cjs@ualberta.ca}
\equalcont{Part of this work was done during an internship at ESA.}

\author*[2]{\fnm{Emmanuel} \sur{Blazquez}}\email{Emmanuel.Blazquez@esa.int}

\affil[1]{\orgdiv{Department of Computing Science}, \orgname{University of Alberta}, \orgaddress{\city{Edmonton}, \country{Canada}}}

\affil[2]{\orgdiv{Advanced Concepts Team}, \orgname{European Space Agency (ESA)}, \orgaddress{\city{Noordwijk}, \country{Netherlands}}}


\abstract{We present a novel approach to system identification (SI) using deep learning techniques.  Focusing on parametric system identification (PSI), we use a supervised learning approach for estimating the parameters of discrete and continuous-time dynamical systems, irrespective of chaos. To accomplish this, we transform collections of state-space trajectory observations into image-like data to retain the state-space topology of trajectories from dynamical systems and train convolutional neural networks to estimate the parameters of dynamical systems from these images. We demonstrate that our approach can learn parameter estimation functions for various dynamical systems, and by using training-time data augmentation, we are able to learn estimation functions whose parameter estimates are robust to changes in the sample fidelity of their inputs. Once trained, these estimation models return parameter estimations for new systems with negligible time and computation costs.}




\maketitle

\section{Introduction}\label{sec:Introduction}
The natural sciences are experiencing a watershed moment with the recent success of sophisticated, data-driven methods for approaching problems that have resisted traditional analytical and optimization-based methods such as AlphaFold for protein structure prediction \cite{Jumper2021}, graph-based neural network models for weather forecasting \cite{Lam2022} and the control of tokamak plasmas using deep reinforcement learning \cite{Degrave2022}. These recent successes have been driven by innovations in large-scale data-driven modeling and rapid advancements in specialized computer hardware that have propelled the rise of deep learning methods alongside domain-specific expertise and techniques from scientific sub-fields. In line with this trend, we present a novel method for solving parametric system identification (PSI) by learning a regression function from return maps to system parameters.

System identification (SI) is the process of modeling and analyzing the behavior of dynamical systems based on observed input and output data\cite{Soderstrom1989}. Accurately characterizing and even predicting the behavior of a dynamical system is an invaluable tool in various areas spanning engineering, physics, biology, and economics. It is particularly valuable in control applications where accurate modeling of a system can result in dramatic performance and robustness improvements \cite{Alvin2003, Bruggeman2007, Ljung2010, Villaverde2014}.
This work focuses on parametric system identification, which is concerned with identifying the underlying parameters of the target system given some parametrized class of dynamical systems believed to contain, or at least closely approximate, the target system \cite{Quaranta2020}.
To understand how and why we solve PSI as a regression problem, it is helpful to place this approach in opposition to the more traditional setting of PSI as an optimization problem to be solved by means of meta-heuristic algorithms such as genetic and evolutionary algorithms\cite{Konnur2005, Tao2007, Wang2011}. Given a parametric class of dynamical systems and data from a target system, in the optimization setting one has a way to compute simulated trajectories given parameter estimates, and the goal is to find parameter estimates which match those of the target system. To accomplish this, one typically combines a loss function which acts as a 'dissimilarity' score between trajectories generated using candidate system parameters and observed trajectories. An optimization scheme is then used to select new candidate parameters for the system. This process is iterated until some computational budget has been exhausted or the loss converges. The observed losses of parameter estimates are then used to select a final estimate, e.g. the parameter estimate with the lowest loss observed in the optimization process. 

One of the challenges of this approach is choosing the loss function that is used to evaluate the quality of a given set of parameter estimates for this solution method. Given equal-length simulated and observed trajectories, one of the most common approaches is to minimize the average error between the pairs of simulated and observed trajectory points \cite{Konnur2005,Tao2007,Ho2010,Modares2010,Wang2011}. This can cause issues when dealing with chaotic systems, which have the characteristic property that trajectories that are initially neighboring in state space can diverge exponentially over time. This means that such a loss function can be highly sensitive to small measurement errors in the initial conditions of the target system. To address this issue, recent work has developed loss functions that compare trajectories as signals at the state space, as opposed to the time-domain level. where the trajectories of chaotic systems tend to take a more structured form \cite{Jafari2014, Mousazadeh2020}. These works have proposed several heuristic loss functions for comparing the similarity of trajectories in state space \cite{Jafari2014, Shekofteh2015, Shekofteh2019, Shekofteh2019a, Mousazadeh2020}, however, there is currently no clear choice as to which state-space loss function is best suited for parameter estimation.

Even working in state space, current methods for PSI solve an optimization problem using an iterative procedure that involves repeatedly simulating trajectories using system parameter estimates, which for continuous-time systems can be a time and computation-consuming process. Worse still, for simple discrete maps, it is often feasible to compute the gradients of the loss function with respect to the system parameters \cite{Mousazadeh2022}, however, a numerical integrator is typically required for continuous-time dynamical systems. This makes evaluating or estimating gradients significantly more challenging. Due in part to this, most previous work has viewed their choice of loss function as a non-differentiable function of the candidate parameters and used various zeroth-order optimization methods such as particle-based methods and genetic algorithms. See \cite{Quaranta2020} for a review of methods based on these computational intelligence approaches. These potentially expensive iterations of simulation and optimization limit the applicability of these methods for real-time settings or processing large collections of systems.

This work explores an alternative approach to PSI for chaotic systems, framing it as a supervised learning problem of mapping state-space data to system parameters. We were initially motivated by the observation that after seeing sufficiently many examples of return maps, a human expert can begin to identify patterns that hint about the parameters of the underlying dynamical systems. Following this intuition, we transform collections of state-space trajectory observations into image-like data to retain the state-space topology of trajectories from dynamical systems and use these collections to train convolutional neural networks to estimate the parameters of dynamical systems from these images. In our approach, the regression loss \textit{is} the parameter estimation error, avoiding the use of heuristic loss functions. Additionally, with almost all of the time and computational cost occurring during data collection and model training, our estimation models return parameter estimations for new samples with negligible time and computation costs at inference time, neatly complementing the drawbacks of optimization-based approaches.

In the remainder of this paper, we formalize the parametric system identification problem as it has conventionally been approached and follow this by presenting our supervised learning problem in \Cref{sec:Background}. We then detail our solution method for this new problem formulation, including generating trajectory datasets, extracting useful features from them, and using them to train a parameter estimation model in \Cref{sec:psi_via_deeplearning}. We present the results of applying our method to both discrete- and continuous-time dynamical systems in \Cref{sec:Experiments} and discuss our findings in \Cref{sec:Discussion}. We conclude by discussing the opportunities and challenges which are introduced with this new approach to analyzing dynamical systems in \Cref{sec:Conclusion}.

\section{Background}\label{sec:Background}
\subsection{Dynamical Systems}\label{subsec:dynamical_systems}
We consider the problem of estimating the parameters of a $p^\text{th}$-order autonomous dynamical system, defined on some state-space $\cX\subseteq \mathbb{R}^p$ with $p\ge1$ by parametric dynamics $F(\cdot~;\mathbf{\mathbf{\theta}}): \mathbb{R}^p \to \mathbb{R}^p$ with $\mathbf{\mathbf{\theta}} \in \Theta \subseteq \mathbb{R}^d$ for some $d\ge1$ by the state equation
\begin{equation}
    \begin{cases}
        \dot{\mathbf{x}}(t) = F(\mathbf{x}(t); \mathbf{\mathbf{\theta}}), \hspace{.0cm} t\in \mathbb{R} &\text{for a continuous flow},\\
        \mathbf{x}_{k+1} = F(\mathbf{x}_k;\mathbf{\mathbf{\theta}}),\hspace{.0cm} k\in \mathbb{Z} &\text{for a discrete map}.
    \end{cases}
    \label{eq:state_eqn}
\end{equation}
Dynamical systems frequently appear in the study of classical mechanics, with the Hamiltonian of some physical system defining the dynamics $F$ on some configuration manifold. For our purposes, the dynamics $F$ can be any function of the state $\mathbf{x}$ and the parameters $\mathbf{\mathbf{\theta}}$ which characterizes the evolution of the state of the system in time according to \cref{eq:state_eqn}.

In this section, we restrict our discussion of parameter estimation to discrete dynamical systems since our approach for handling continuous-time systems is to use Poincaré sections and their respective Poincaré maps to represent them with discrete-time systems for analysis\cite{Parker1991}. 

\subsubsection{Poincaré Maps}\label{para:poincare}
A Poincaré map is defined by a dynamical flow along with an oriented hypersurface in state space, referred to as a Poincaré section. Some references remove the requirement that the section is oriented, resulting in a so-called `two-sided' Poincaré map. This choice does not have significant consequences for our purposes. A Poincaré map is the unique map associated with the original dynamical system, which takes a point $P$ in the Poincaré section to the next point $P'$ in the Poincaré section where the flow of the original dynamical system, originating from $P$ crosses the oriented section in the same direction as at $P$. 
For this reason, Poincaré maps are sometimes called `first recurrence', or `return maps'. In this work, when we refer to return maps, we refer either to a collection of state-space observations from a discrete map or from the discrete map induced by a continuous time dynamical system along with a particular choice of Poincaré section. Poincaré maps allow us to present a unified method for parametric system identification on both discrete maps and continuous time dynamical systems.

\subsection{Parametric System Estimation}\label{subsec:PSI}
 In the parametric system identification problem, we are given the parametric form of $F$ and are tasked with constructing an estimate $\hat{\mathbf{\mathbf{\theta}}}^*$ on the basis of a collection of $n$ observed trajectories $\mathbf{T} = (\mathbf{\tau}_1, \mathbf{\tau}_2,\ldots, \mathbf{\tau}_n)$ where each trajectory $\tau_i = (\mathbf{x}^i_1, \mathbf{x}^i_2, \ldots)$ is an indexed collection of elements from $\cX$ which follows $\mathbf{x}^i_{k+1} = F(\mathbf{x}^i_k;\mathbf{\mathbf{\theta}}^*)$. Most previous work on PSI considers the case when $n=1$, but is straightforward to generalize for the case of multiple sample trajectories.

The optimization approach to estimating $\mathbf{\mathbf{\theta}}^*$ from a collection of observed trajectories of the target system, $\mathbf{T}$ is by solving an optimization problem of the form
\begin{equation}\label{eqn:optim}
    \hat{\mathbf{\mathbf{\theta}}}^* = \argmin_{\mathbf{\mathbf{\theta}} \in \Theta} ~\ell(\mathbf{\mathbf{\theta}}, \mathbf{T}),
\end{equation}
where $\ell: \Theta \times \cX^{n\times m} \rightarrow \mathbb{R}^+$
is a loss function that maps parameters estimates to positive real numbers, given observed trajectories $\mathbf{T}$, and where we assume for simplicity that each of the $n$ observed trajectories in $\mathbf{T}$ consist of $m$ points. The value of $\mathbf{\mathbf{\theta}}^*$ influences $\mathbf{T}$ through the dynamics $F(\cd; \theta^*)$ imposed on the trajectories. For this reason, previous work has typically defined $\ell$ by composing a loss function $f$ that compares sets of trajectories with a numerical solver that is used to produce simulated trajectories corresponding to a parameter estimate. Concretely, to evaluate the loss of a parameter estimate $\hat{\mathbf{\theta}}^*$ against the true parameter $\mathbf{\theta}^*$, $\ell(\hat{\mathbf{\theta}}^*, \mathbf{\theta}^*)$ practitioners use a numerical solver to create a collection of trajectories $\mathbf{T}(\hat{\mathbf{\theta}}^*) = (\hat{\mathbf{\tau}}_1, \hat{\mathbf{\tau}}_1, \ldots, \hat{\mathbf{\tau}}_n)$ with $\mathbf{\tau}_i = (\hat{\mathbf{x}}^i_1, \hat{\mathbf{x}}^i_2, \ldots, \hat{\mathbf{x}}^i_m), \hat{\mathbf{x}}^i_k \in \cX$ and $\hat{\mathbf{x}}^i_{k+1} = F(\hat{\mathbf{x}}^i_k;\hat{\mathbf{\theta}}^*)$. They then evaluate a loss that assigns values to pairs of collections of trajectories $f(\mathbf{T}', \mathbf{T})$. This composition of a numerical solver and loss function over sets of trajectories defines an implicit loss function on $\Theta$ given a collection of observed trajectories $\mathbf{T}$,

\begin{equation}\label{eqn:implicit_opt}
    \ell(\hat{\mathbf{\theta}}, \mathbf{T}) \ceq f(\mathbf{T}(\hat{\mathbf{\theta}}), \mathbf{T}).
\end{equation}

The method proposed in \cite{Mousazadeh2022} does not follow this description, though as mentioned earlier, this method is not applicable when numerical methods are required to approximate trajectories, e.g., when they correspond to a Poincaré map. 


\subsubsection{State-space Representations of Trajectories}\label{subsec:state_space_repr}
Parameter estimation is especially challenging when dealing with dynamical systems which exhibit chaotic behavior. This is because of the characteristic sensitivity of these systems to their initial conditions. Even if the exact parameters of a chaotic system are known, any measurement error in the system's initial conditions can result in a simulated trajectory that diverges exponentially from the target observation set over time. Due to these issues, choices of $f(\hat{\mathbf{T}}(\hat{\mathbf{\theta}}), \mathbf{T})$ which are constructed around pairwise differences of $\hat{\mathbf{x}}^i_k$ and $\mathbf{x}^i_k$ in the time-domain are often poorly-conditioned with respect to $\hat{\mathbf{\theta}}$ as well as any measurement error in the initial conditions $\mathbf{x}^1_1, \mathbf{x}^2_1, \ldots, \mathbf{x}^m_1$ of the target system. A choice of $f$ which appears frequently in the literature is the temporal mean squared error (MSE)
\begin{equation}\label{eqn:time_MSE}
    f_{\text{MSE}}(\hat{\mathbf{T}}(\hat{\mathbf{\theta}}), \mathbf{T}) \ceq \dfrac{1}{mn} \sum_{i=1}^n\sum_{k=1}^m \| \hat{\mathbf{x}}^i_k - \mathbf{x}^i_k \|^2.
\end{equation}

To address these issues, recent work such as \cite{Jafari2014, Mousazadeh2020} has investigated the use of state-space representations of trajectories to compare simulated trajectories to the target data. The basic idea of this method is outlined by Jafari et al. \cite{Jafari2014}. While chaotic systems appear to have highly disordered behavior when considering trajectories in the time domain, they are known to have a more structured topology in state-space, which for our purposes refers to considering trajectories as unordered sets of points in state-space as opposed to time-indexed points. Chaotic attractors manifest this phenomenon, wherein collections of dynamical system trajectories that appear stochastic in the time domain are constrained to a low-dimensional manifold in state space. This property of chaotic systems makes using state-space representations of trajectories from dynamical systems appealing for designing the function $f$ in the right-hand side of \cref{eqn:implicit_opt}.

The main contribution of \cite{Jafari2014} was the construction of a loss function between \textit{return maps}. Their loss function considers $\mathbf{T}(\hat{\mathbf{\theta}})$ and $\mathbf{T}$ as structureless sets of points in $\cX$, and measures the average of the Euclidean distances between each point in the simulated data with those of its nearest neighbor in the target data, and vice versa. 

\subsection{Supervised Machine Learning for Parametric System Identification}\label{subsec:ML_for_PSI}
So far, we have only discussed the optimization approach to parametric system identification (\cref{eqn:optim}). Our novel solution for PSI instead frames parameter estimation as a supervised machine-learning problem (see e.g., \cite{Murphy2012}). In our approach, we aim to identify a single function that maps from observations of trajectories from a discrete dynamical system to an estimate of the system's parameter $g:\cX^{n \times m } \rightarrow \mathbb{R}^d$, where $g\in \mathcal{F}$ is an element of some function class $\mathcal{F}$. We select $g$ through a data-driven optimization process that attempts to minimize the parameter estimation error of $g$, assuming access to a large dataset of input-output pairs $\cD=(\mathbf{T}_i, \mathbf{\theta}_i)_{i=1}^N$ with a \textit{wide range} of values $\theta_i \in \Theta$.
The $i^{\text{th}}$ pair in this set consists of a collection of trajectories $\mathbf{T}_i$ along with the parameter, $\mathbf{\theta}_i$, which generated them. Concretely, given a parameterized class of dynamical systems, we use real or simulated observations from systems with known parameters to construct a dataset, $\cD$. We then use $\cD$ to estimate an optimal parameter estimation function, $g^*$, defined to minimize the mean squared prediction error, or another suitable choice of loss function, over some joint probability distribution $\mathbb{P}$ of possible values of $\mathbf{\theta}^*$ and the resulting observations $\mathbf{T}(\mathbf{\theta}^*)$:
\begin{equation}\label{eqn:ERM}
    g^* = \argmin_{g \in \mathcal{F}}~ \mathbb{E}_{(\mathbf{\theta},\mathbf{T}) \sim \mathbb{P}}\sbr{\|\mathbf{\theta} - g(\mathbf{T}(\mathbf{\theta}))\|^2}.
\end{equation}

For example, $\mathbb{P}$ could be the joint distribution corresponding to $\theta$ drawn uniformly over a bounded subset of $\mathbb{R}^d$ with trajectories $\mathbf{T}(\mathbf{\theta})$ generated from $\mathbf{\theta}$ in some consistent but possibly random fashion, e.g., applying a deterministic numerical integration software to randomly sampled initial states.

Given an estimate of the minimizer of \cref{eqn:ERM}, $\hat{g}^*$, and a new collection of trajectories $\mathbf{T}(\mathbf{\theta}')$ corresponding to parameter $\mathbf{\theta}'$, we take $\hat{g}^*(\mathbf{T}(\mathbf{\theta}'))$ to be our estimate $\hat{\mathbf{\theta}}'$. We have purposefully left the form of the function class $\mathcal{F}$ vague in this section. These details are clarified in \Cref{subsec:ML_for_PSI} when we outline our supervised machine learning approach.

\section{Parameter Estimation via Deep Learning on Return Maps}\label{sec:psi_via_deeplearning}

Estimating system parameters directly from observed trajectories is a challenging problem for the same reasons outlined in \Cref{subsec:state_space_repr}. Drawing from the lessons of earlier work on PSI for chaotic systems, rather than working with trajectories as time-series data, we restrict our attention to functions $g$ which only consider the state-space information of the trajectory collections $\mathbf{T}_i$ that are passed as input. Further, we transform this data into a form that allows us to make use of existing CNN architectures to construct a parameter estimation function. Specifically, we discretize regions of state space into pixels and use this discretization scheme to transform return maps into coarse, single-channel images, and use these as the input `features' for our models.

\begin{figure}[ht!]
    \begin{center}
        \includegraphics[width=1\columnwidth]{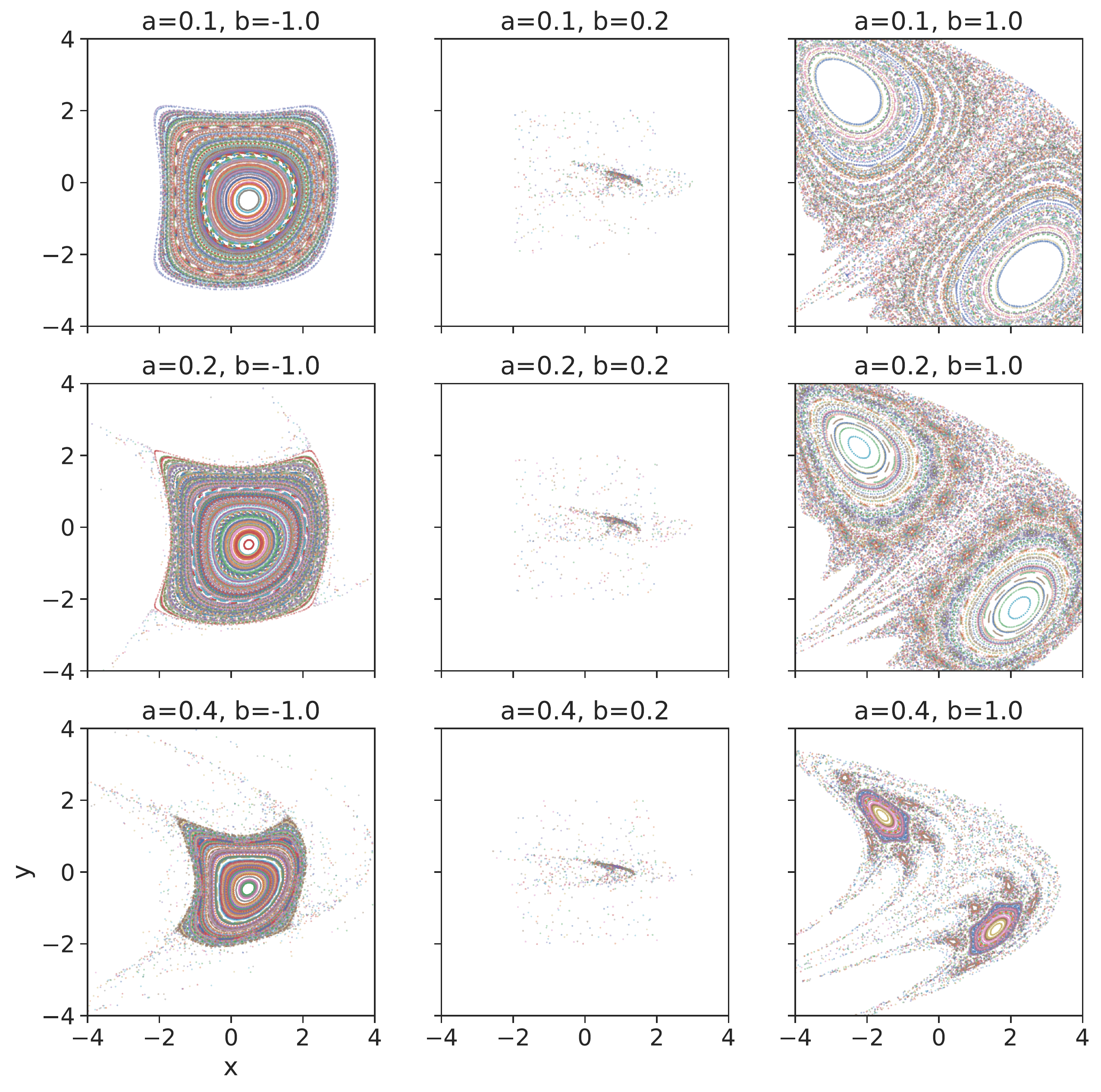}
    \end{center}
    \caption{Return maps generated by sampling states from the uniform distribution over the region $x,y \in [-2,2]\times[-2,2]$ and plotting 250 iterations of the Hénon Map (\cref{eqn:henon}) from each initial state. The different colors correspond to trajectories for different initial states.}\label{fig:henon_return_map}
\end{figure}


\Cref{fig:henon_return_map} conveys the central intuition for our work. Different parameter values for the Hénon map result in visually distinct return maps. This observation suggests that it might be possible to learn a regression function from these images to parameter estimates. Our experiments show that it is possible to learn such a regression function from data.

The main contribution of this work is the demonstration that our method of predicting system parameters from images of return maps is viable, introducing the possibility of further applications of deep learning on return maps. For simplicity, we consider dynamical systems with two-dimensional return maps in this work. However, conceptually, our method generalizes to higher dimensions at the cost of the requirement of additional computation and data requirements commensurate with the curse of dimensionality \cite{Murphy2012}.

\subsection{Dataset Generation}\label{subsec:data_gen}
Developing a supervised machine learning model for predicting a system's parameters necessitates a dataset to train the model. Dataset generation is straightforward after specifying a parameterized dynamical system and the range of parameter values of interest. To ensure that the trained model can perform well on the entire range of parameter values, we first selected a large collection of evenly spaced parameter values. For the $i^{\text{th}}$ parameter value $\mathbf{\theta}_i$, we sampled $n$ initial states conditions and iterated the dynamical map $m-1$ times in the discrete case, or made use of Heyoka numerical integration software \cite{Biscani2021} to evaluate a total of $m$ crossing points for a specified crossing section. The result of this generation process was a collection of input-output pairs, $\mathbf{T}_i$ where the `input' is $\mathbf{\theta}_i$, and the `output' is a collection of discrete trajectories corresponding to the $N_\text{traj}$ trajectories of length $m$. In the case of dynamical flows, it can be simpler to propagate the system forward until a fixed system time. This can lead to the trajectories starting from the different initial states having different numbers of crossings through the section. This detail does not introduce any issues for our method, but it is a consideration in practice.

\subsection{Feature Processing}\label{subsec:feature_proc}
\begin{figure}[ht!]
    \begin{center}
        \includegraphics[width=1\columnwidth]{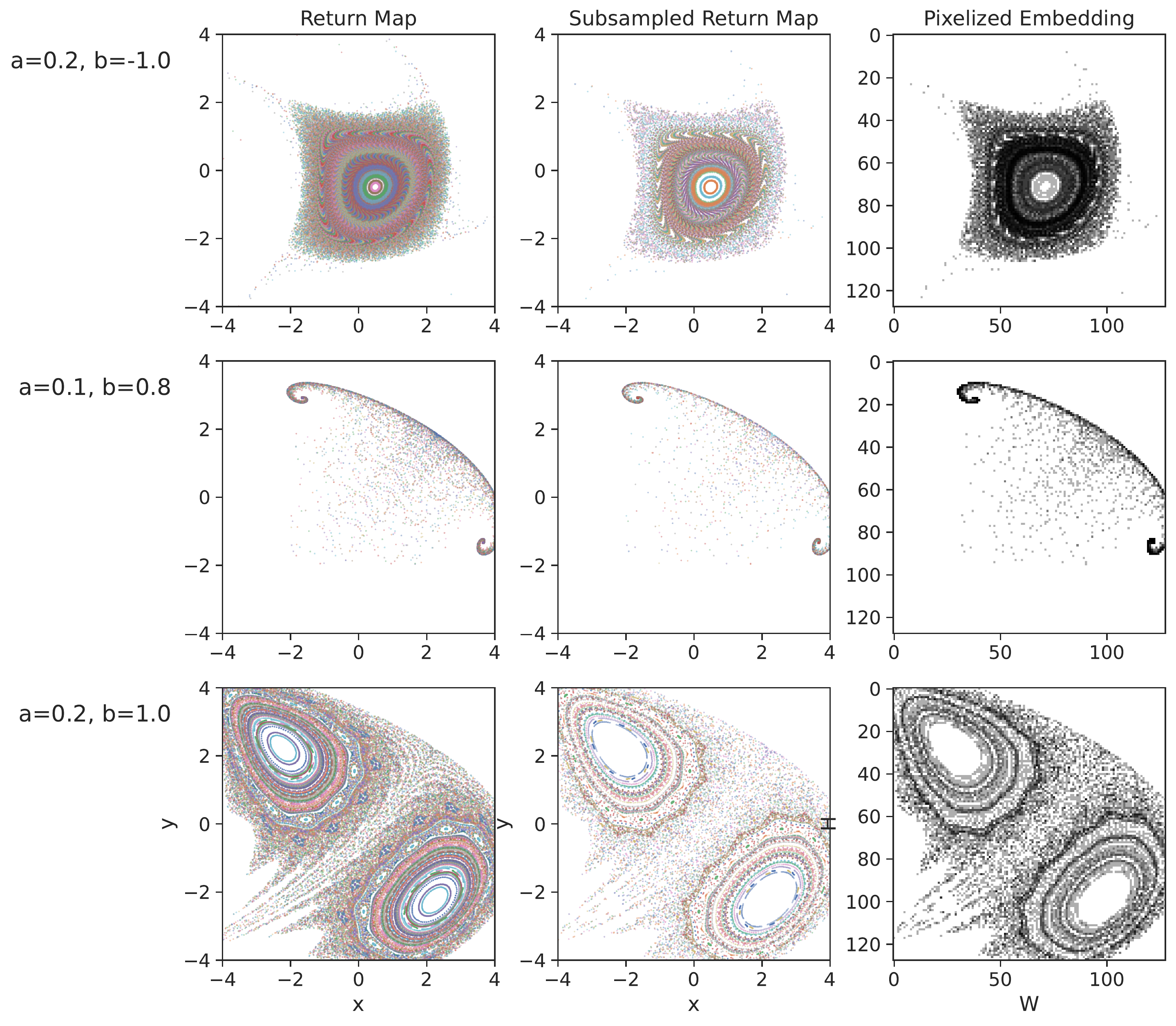}
    \end{center}
    \caption{Transforming subsampled return maps into images. Each row shows the process of processing a return map for input to the model. The left column shows return maps sampled in the same fashion as in \Cref{fig:henon_return_map}. The middle column shows the output of our random data augmentation scheme, i.e. taking a random subset of the trajectories and truncating them to a random length. The right-most column shows examples of the single-channel images fed input to the model. The images are obtained by discretizing a region of state-space, here the region $ \mathbf{x} \in [-4,4]^2$, into pixels and darkening each pixel following \cref{eqn:pixel_shading}.}\label{fig:featurization}
\end{figure}
As discussed in the previous section, there are inherent benefits to considering state-space representations of trajectories. For each input-output pair $(\mathbf{T}_i, \mathbf{\theta}_i)_{i=1}^N$ in our dataset, we take the trajectories $\mathbf{T}_i$ and `flatten' them into a state-space representation, essentially overlaying coarse scatter-plots of each trajectory in state-space. We define this transformation by selecting a state-space region and splitting it into `pixels', or a higher-dimensional discretization of the space in a more general setting, e.g. `voxels' in three dimensions. Our experiments were performed with an axis-aligned uniform grid of $128 \times 128$ pixels. Given the $i^\text{th}$ collection of trajectories $\mathbf{T}_i$, the pixels were `shaded' by defining the value of the pixel in the $(h,w)\in \{0,1,...127\}^2$ position in the $i^{\text{th}}$ sample by
\begin{equation}\label{eqn:pixel_shading}
    \mathbf{P}_{i,h,w} \ceq \alpha^{n_{i,h,w}},
\end{equation}
where $n_{i,h,w}$ is the number of points in $\mathbf{T}_i$ which lie in the region of space ascribed to the pixel at position $(h,w)$ and $\alpha \in (0,1]$ is a transformation parameter which determines the exponential base that the pixels `darken' with. We will use $\mathbf{P}_i$ to refer to the collection of pixels $\mathbf{P}_{i,h,w}$, or 'pixelized return map'. The input-output pairs after this transformation are then $(\mathbf{\theta}_i, \mathbf{P}_i)_{i=1}^N$. The result of this process is shown in the rightmost column of \Cref{fig:featurization}. We performed all experiments with $\alpha=7/10$. 

\subsubsection{Data Augmentation}\label{subsubsec:data_aug}
We experimented with an additional step during model training before creating each pixelized return map. Data augmentation \cite{Shorten2019} is a common practice in training machine learning models in which carefully chosen random modifications are made to training samples to improve the generalizability of models to new data. In short, data augmentation can reduce the sensitivity of the estimation performance of trained models to certain changes to their input. An example of this method in image recognition tasks is making small random changes to the cropping of input photos. The rationale is that, for example, the subject of a photo with small differences in cropping is the same, and we would like our models to respect this invariance. An additional benefit of data augmentation is that it acts as a regularization mechanism that reduces the tendency of trained models to overfit to patterns in the training dataset at the expense of worse performance on new data. In our setting, we performed data augmentation to make our model robust to changes in the number and length of the trajectories we flattened to form the images. 
Specifically, given the $i^\text{th}$ input-output pair, we first sampled $N_{\text{traj}} \sim \text{Uniform}(10,n)$. We then selected $N_{\text{traj}}$ of the $n$ trajectories in $\mathbf{T}_i$ uniformly at random. Next, we sampled $N_{\text{steps}} \sim \text{Uniform}(10,m)$, and for each of the $N_{\text{traj}}$ trajectories we took only the first $N_{\text{steps}}$ points. 
The resulting collection of $N_{\text{traj}}$, length $N_{\text{steps}}$ trajectories was then used to create a pixelized return map as discussed in \Cref{subsec:feature_proc}. \Cref{fig:featurization} shows the process of creating pixelized return maps with this data augmentation method.
The choice of 10 as the lower limit on the number of trajectories and steps in the augmentation step is another hyperparameter. Although we do not explore other choices for this value in this work, one should consider that choosing a value too close to $m$ or $n$ limits the diversity of the training data; this may reduce the regularization effect. On the other hand, training the model on samples with too few trajectories or with too few steps may result in a trained model which fails to generalize to more densely-sampled return maps.

\subsection{Parameter Estimation Model}\label{subsec:ML_model}
\begin{figure*}[ht!]
    \begin{center}
        \includegraphics[width=.85\textwidth]{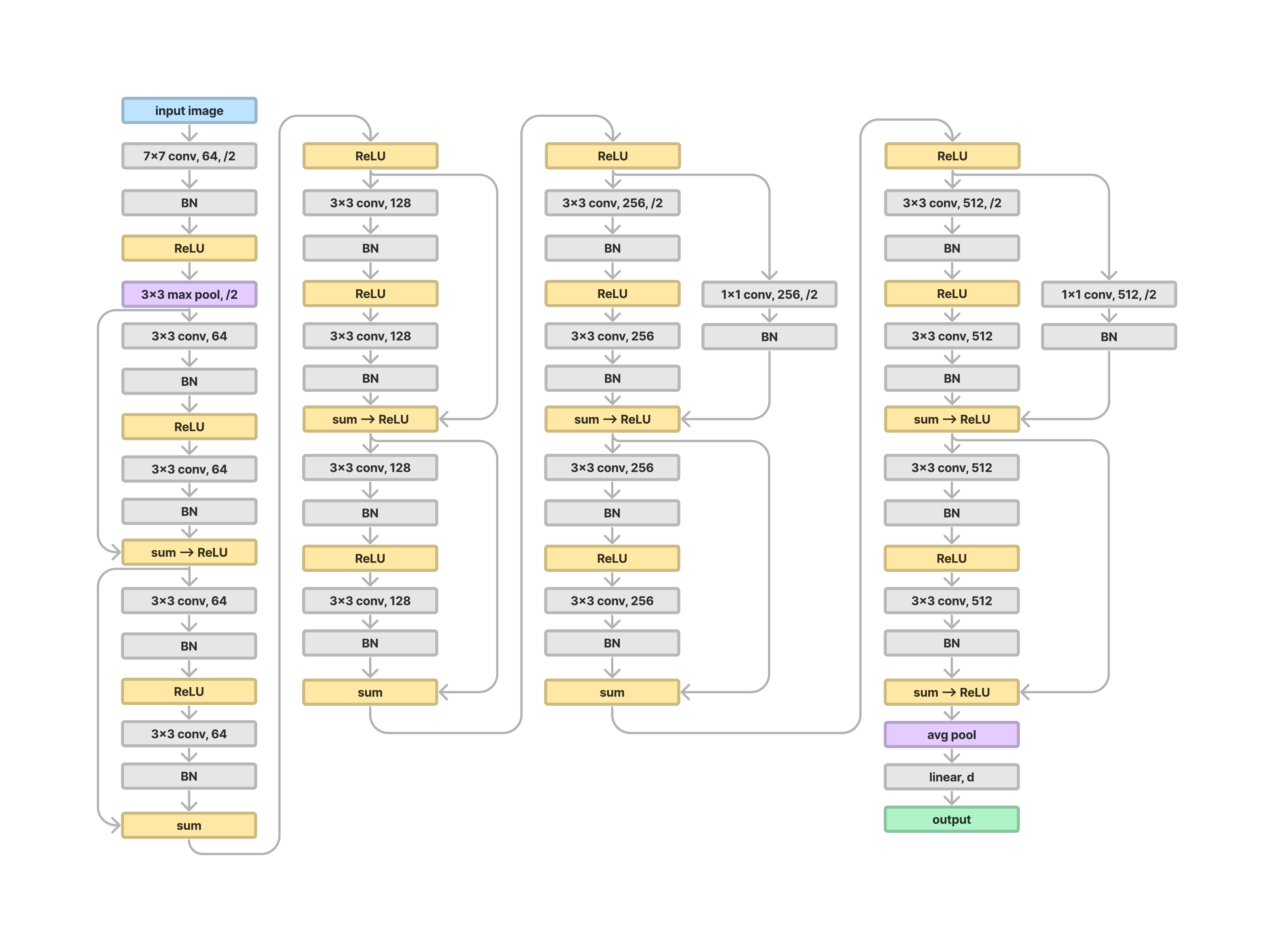}
    \end{center}
    \caption{A network architecture diagram for the ResNet18 neural network \cite{He2016}. The convolutional (conv), batch-norm (BN) \cite{Ioffe2015a}, and linear layers contain `learnable' parameters, which we optimize during the model training process using stochastic gradient descent. The network's name comes from the fact that each input image passes through 17 convolutional layers and one linear layer. We used the PyTorch \cite{Paszke2017} TorchVision \cite{TorchVisionMaintainersAndContributors2016} library's implementation of ResNet18, only modifying the first convolutional layer to take single channel images as input.}\label{fig:resnet18}
\end{figure*}   
Our experiments were all performed with a ResNet18 neural network \cite{He2016}, a small model from a widely used family of deep-learning models which feature convolutional layers and residual connections designed for computer-vision tasks. The structure of the network is depicted in \Cref{fig:resnet18}. We selected this architecture as the model's residual connections and batch-normalization layers \cite{Ioffe2015a} lead to relatively easy training of the model with current methods. In the context of the formalism that we introduced in \Cref{subsec:ML_for_PSI}, we consider estimation functions $g\in \mathcal{F}$ which are the composition of our transformation from collections of trajectories to images of return maps, with a ResNet18 neural network with learnable parameters. Selecting a specific function $g$ amounts to setting the learnable parameters of the ResNet18, and possibly the parameters which describe the transformation from trajectories to images.

\subsubsection{Model Training}\label{subsubsec:model_training}
Given a dataset $\cD$ of $N$ input-output pairs, we partitioned the pairs into training ($\cD_\text{train}$), validation ($\cD_\text{validation}$) and testing ($\cD_\text{test}$) datasets containing respectively 65\%, 15\% and 20\% of the pairs in $\cD$. This split was performed in a manner that ensured that each partition contained a representative mix of samples from the full range of parameter values $\Theta$. 

As is standard in training supervised machine learning models, $\cD_\text{train}$ and $\cD_\text{validation}$ were used to optimize the parameters of each model. Specifically, $\cD_\text{train}$ was used to optimize the learnable parameters of the ResNet18 model using stochastic mini-batch gradient descent with the Adam optimizer \cite{Kingma2017} as our optimization routine. We used default parameters for the optimizer other than the learning rate and value of the weight decay parameter. When using data augmentation, we performed fresh random augmentation on each pass over the training set, i.e., the model essentially never saw the same sample twice in the training routine. When applicable, the validation set $\cD_\text{validation}$ had random data augmentation applied only once at the start of training. We fixed the validation set augmentation to provide a consistent set to evaluate the performance of each model, reducing the variance of the validation error. We used the model's error on $\cD_\text{validation}$ to optimize the hyper-parameters of our optimization routine. We focused on tuning the Adam optimizer's learning rate and weight decay parameters, as well as the number of training samples used to evaluate each step of the optimizer (commonly referred to as the training batch size). After model training, we opted to select the model version with the smallest validation loss observed during the training process. The resulting model's performance was then evaluated on $\cD_\text{test}$, providing an unbiased estimation of model performance as the test set consists of samples not used in the model training or model selection process.

\subsection{Loss Function}\label{subsec:loss_function}
By framing parametric system identification as a supervised machine learning problem, we can develop an estimation model that directly optimizes a loss function on the estimation error. Still, the appropriate loss function to optimize depends on the intended purpose of the parameter estimates. For example, simply choosing the mean squared error may be sufficient for systems with a single parameter. However, for systems with parameters $\mathbf{\theta} \in \mathbb{R}^d$ for $d>1$, we may require greater precision when estimating some coordinates of $\mathbf{\theta}$ than others. One way to encode this information is in the loss function used during model training. For example, in \Cref{subsec:henon_map} we use a weighted loss function to ensure that the relative errors of our model's parameter estimates are roughly equal for both Hénon map parameters.

\begin{figure*}[ht!]
    \begin{center}
        \includegraphics[width=.85\textwidth]{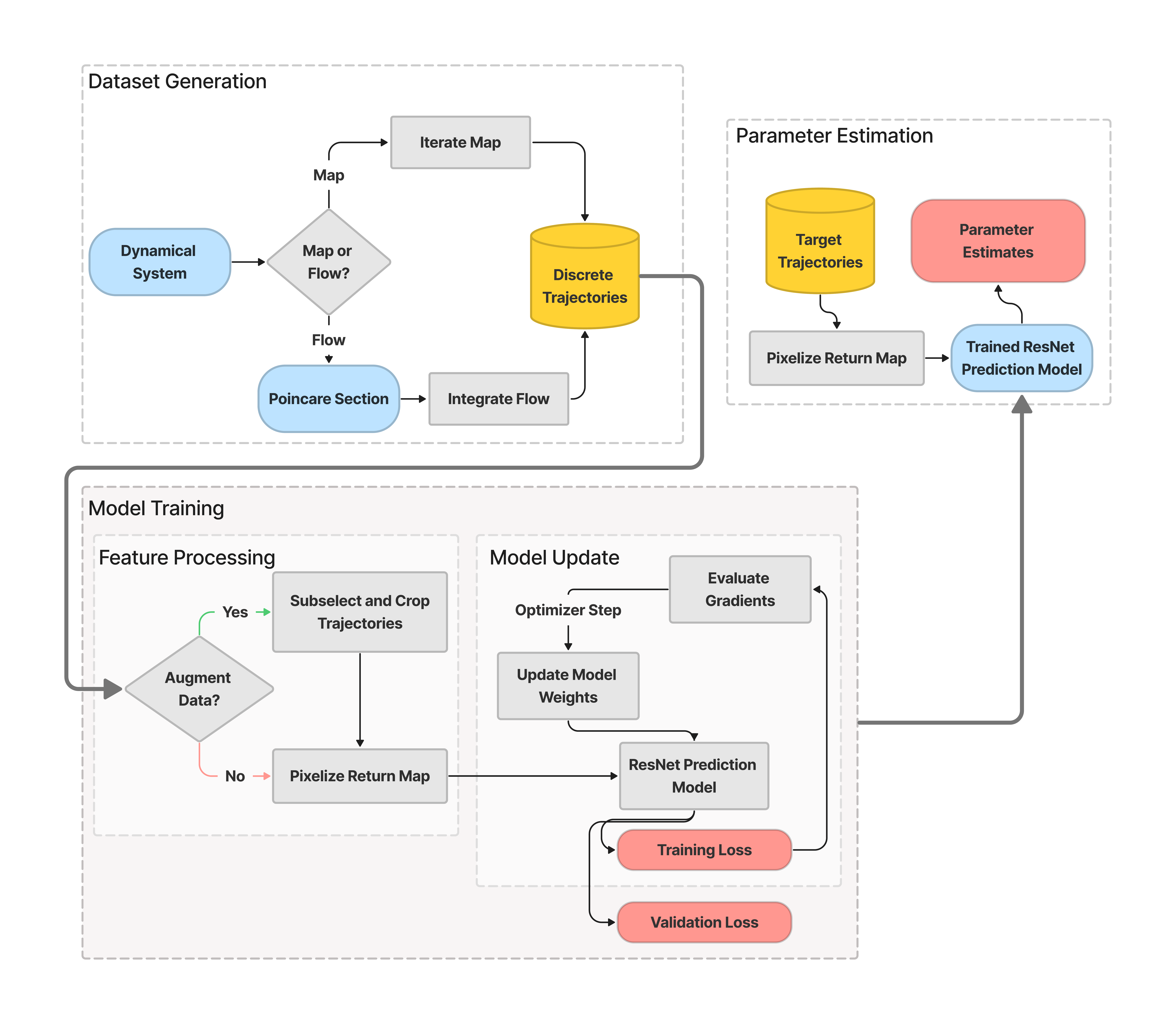}
        \caption{Schematic diagram showing the process of generating training data, training the parameter prediction model, and finally using the trained model for parameter estimation.}\label{fig:diagram}
    \end{center}
\end{figure*}

\Cref{fig:diagram} outlines the entire process of creating a dataset from a dynamical system, training a supervised machine learning model, and deploying the model for parameter estimation.

\section{Experiments}\label{sec:Experiments}

We trained parameter estimation models on continuous- and discrete-time dynamical systems and performed experiments to understand the sensitivity of our method to the amount of available training data, as well as the impact of our data augmentation process. We show results for two systems in this work. In order to build intuition for our proposed method, we begin with the simpler case of parameter estimation for a discrete map before showing experiments for a continuous-time dynamical system. In what follows we will refer to models that are trained with or without augmented data as \textit{augmented} and \textit{non-augmented} models respectively.

All of our experiments were carried out using a single NVIDIA\textsuperscript{\textregistered} GeForce\textsuperscript{\textregistered} RTX 2080 Ti GPU and two Intel\textsuperscript{\textregistered} Xeon\textsuperscript{\textregistered} E5-2650L v4 CPUs with a clock rate of 1.70GHz and 251GB of RAM.

\subsection{Hénon Map}\label{subsec:henon_map}
In our first experiment, we estimate 
the parameters of the Hénon Map \cite{Henon1976}, a well-studied discrete dynamical system that maps pairs of points $(x_k, y_k) \in \mathbb{R}^2$ to $(x_{k+1}, y_{k+1})$ according to
\begin{equation}\label{eqn:henon}
    \begin{cases}
        x_{k+1}=1-ax_k^2 + y_k\\
        y_{k+1}=bx_k,
    \end{cases}6
\end{equation} 
where $a,b \in \mathbb{R}$ are the parameters which characterize the system.

For our experiments, we considered the Hénon map parameters $a\in[0.05,.45], b\in[-1.1,1.1]$. We constructed a database of input-output pairs by taking a large number of $a,b$ points spaced evenly in $[0.05,.45]\times[-1.1,1.1]$. For the $i^\text{th}$ $a,b$ pair, we generated trajectories by applying the Hénon Map with parameters $a,b$ $250$ times over a uniform grid of $225(=15^2)$ initial $x,y$ points over the region $[-4,4]\times [-4,4]$. With this collection of parameters and trajectories $\cD$, we processed features as outlined in \Cref{subsec:feature_proc} to obtain image-like data, which we then used to train a ResNet18 neural network to make parameter estimates in $\mathbb{R}^2$. Given the different scales of the $a$ and $b$ parameters in this task, we chose to measure our parameter prediction performance according to a weighted mean squared error loss
\begin{equation}
    \ell\rbr{(\hat{a}, \hat{b}), (a, b)} = \dfrac{1}{2}\rbr{\sigma_a^2 (\hat{a} - a)^2 + \sigma_b^2 (\hat{b} - b)^2},
\end{equation}
where $\sigma_a, \sigma_b=(25, 4.\overline{54})$ were chosen so that $\sigma_a \cd (a-a_\text{mid})$ and $\sigma_b \cd (b-b_\text{mid})$ take values in the interval $[-5,5]$, where $(a_\text{mid}, b_\text{mid})=(-.25, 0)$ is the center of the parameter region $[0.05,.45] \times [-1.1,1.1]$

\subsubsection{Results}\label{subsubsec:henon_results}
\begin{figure}[ht!]
    \begin{center}
        \includegraphics[width=\columnwidth]{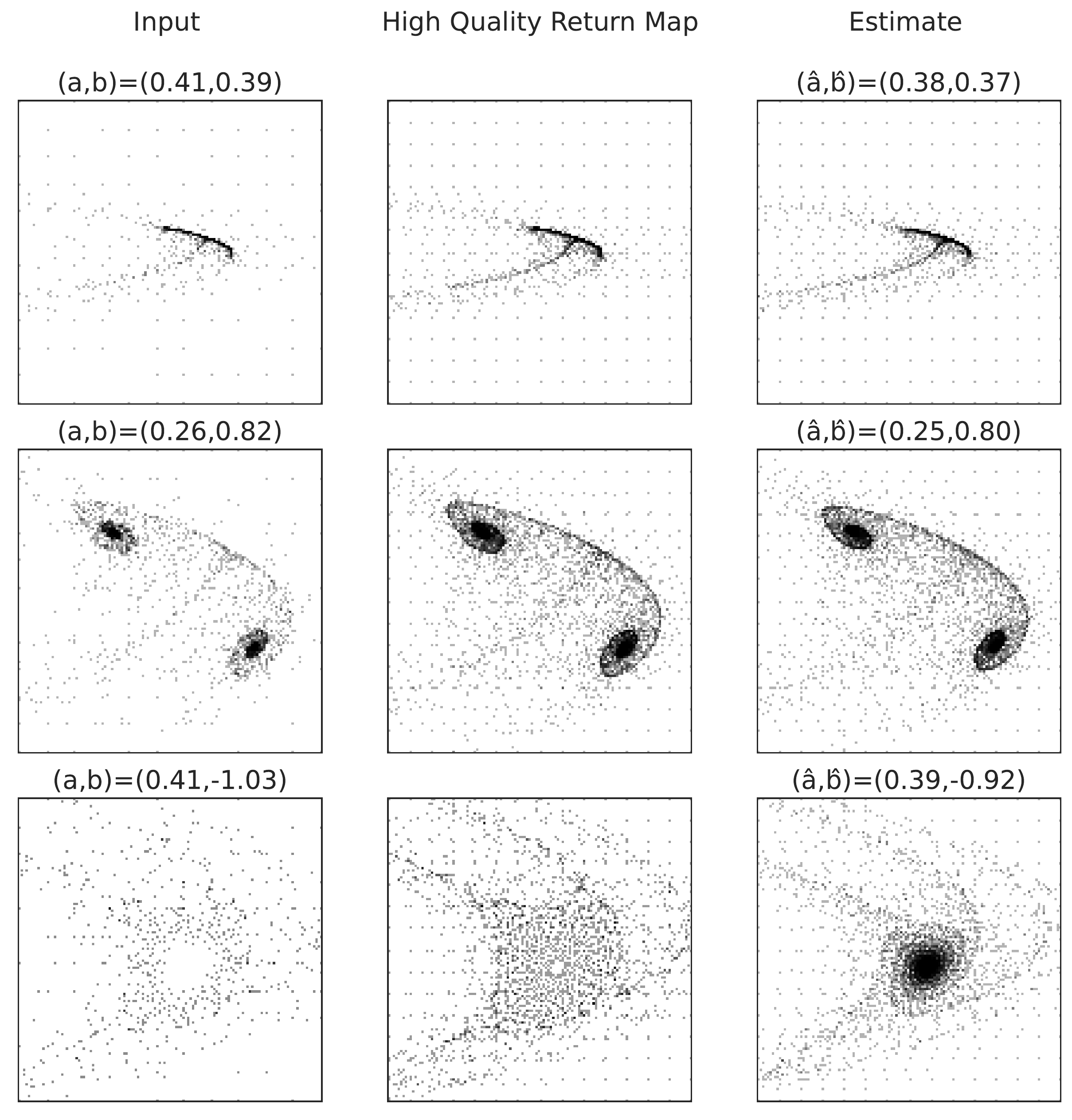}
        \caption{Inputs and outputs from a trained parameter estimation model for the Hénon map taking randomly augmented return map images as input. The left column shows the input to the network, as well as the ground truth values of $a,b$. The center column shows high-quality return map images, not input to the network, corresponding to the system's parameters, obtained by iterating the map for a uniform grid of initial conditions using actual values of $a,b$. The right column shows the model outputs and a return map obtained in the same fashion as the middle column but using the model's \textit{estimated values} $\hat{a},\hat{b}$.}\label{fig:henon_predictions}
    \end{center}
\end{figure}

\Cref{fig:henon_predictions} shows input and output examples from a test set for a trained parameter estimation model. As well as displaying the output parameter estimates of the network, the figure shows return maps that correspond to these parameter estimates, providing a qualitative assessment of our model's estimation accuracy. The inputs to the estimation model were noisily sampled in the same fashion as in our data augmentation process (\Cref{subsubsec:data_aug}). We also show more refined return maps, generated without this noisy sampling, that correspond to the true parameter values for a more direct comparison with the return maps corresponding to the estimation model's outputs. For a perfect estimation model, the return maps in the middle and right columns would be essentially identical. The example in the third row of \Cref{fig:henon_predictions} illustrates the challenge of estimating parameters from the sparsely sampled return maps that our data augmentation process can create. While the `ground truth' return map and the return map corresponding to the parameter estimates visually differ near the center of the image but are otherwise similar. The input to the network lacks samples in this central region, which makes the model's estimate plausible given its input.

\begin{figure}[ht!]
    \begin{center}
        \includegraphics[width=1\columnwidth]{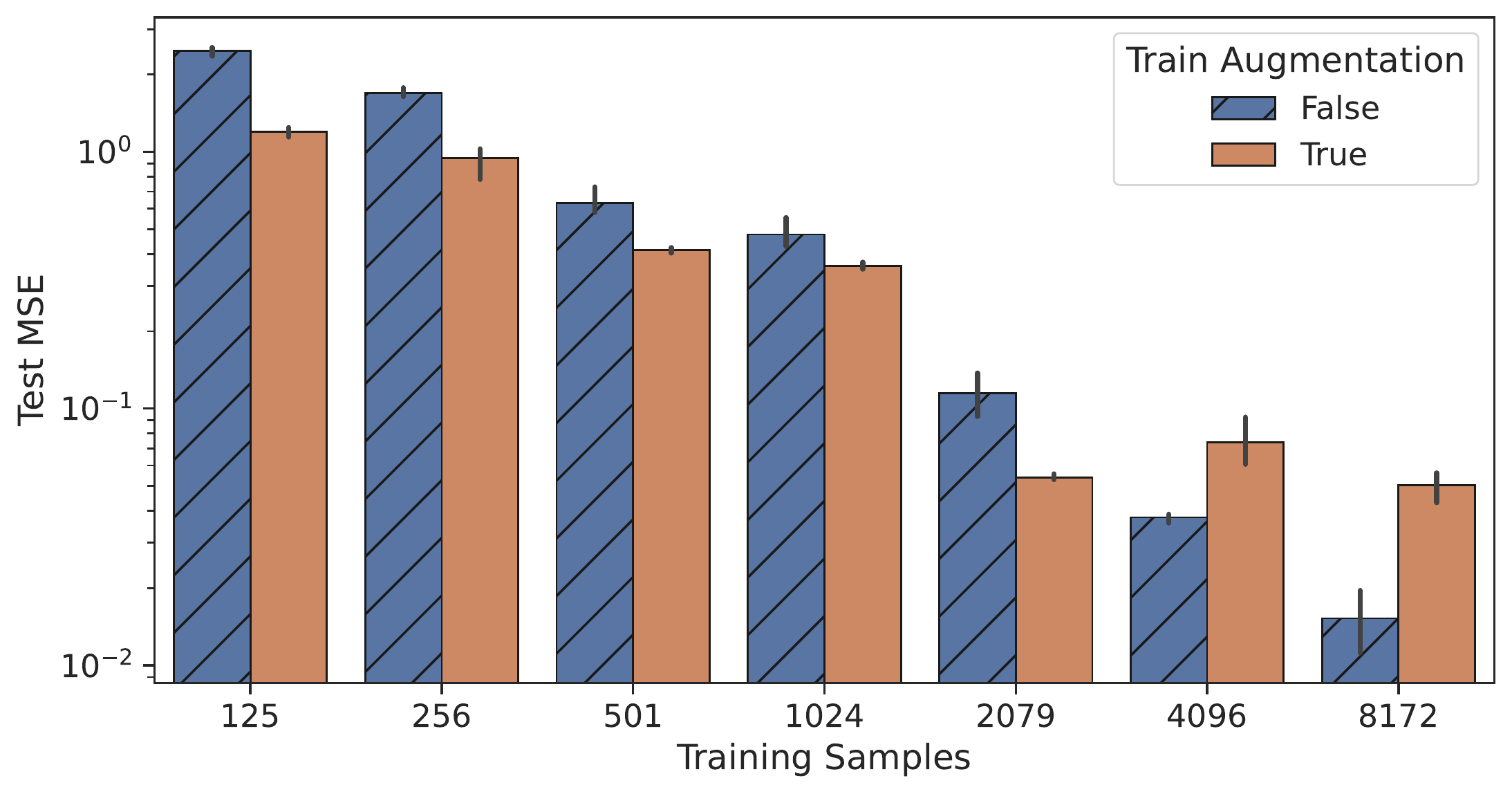}
    \end{center}
    \caption{Test mean squared error of augmented and non-augmented models trained on datasets of increasing numbers of sample pairs from the Hénon map. Each model was trained for 25,000 optimization steps with the same configuration of the Adam optimizer, i.e. the same learning rate, weight decay, and batch size parameters. The five model versions with the lowest validation error observed during training were selected for each dataset size. We then evaluated these models on a test set. We evaluated model errors on inputs that matched their respective training distributions. I.e., we only applied data augmentation to the test samples for augmented models.  Error bars represent bootstrapped 95\% confidence intervals for models trained with three random seeds.
    }
    \label{fig:henon_data_aug_barplot}
\end{figure}


\Cref{fig:henon_data_aug_barplot} summarizes the results of an experiment investigating the impact of training dataset size and our data augmentation process on parameter estimation error on the Hénon map. We emphasize two aspects of this figure, the first of which is that the test error of our estimation models decreases in an approximately power-law fashion with the size of the training set. 
A second aspect of \Cref{fig:henon_data_aug_barplot} is that the augmented models seem to have lower test error than non-augmented models on smaller numbers of training samples until this trend reverses on larger training datasets.

\begin{figure}[ht!]
    \begin{center}
        \includegraphics[width=1\columnwidth]{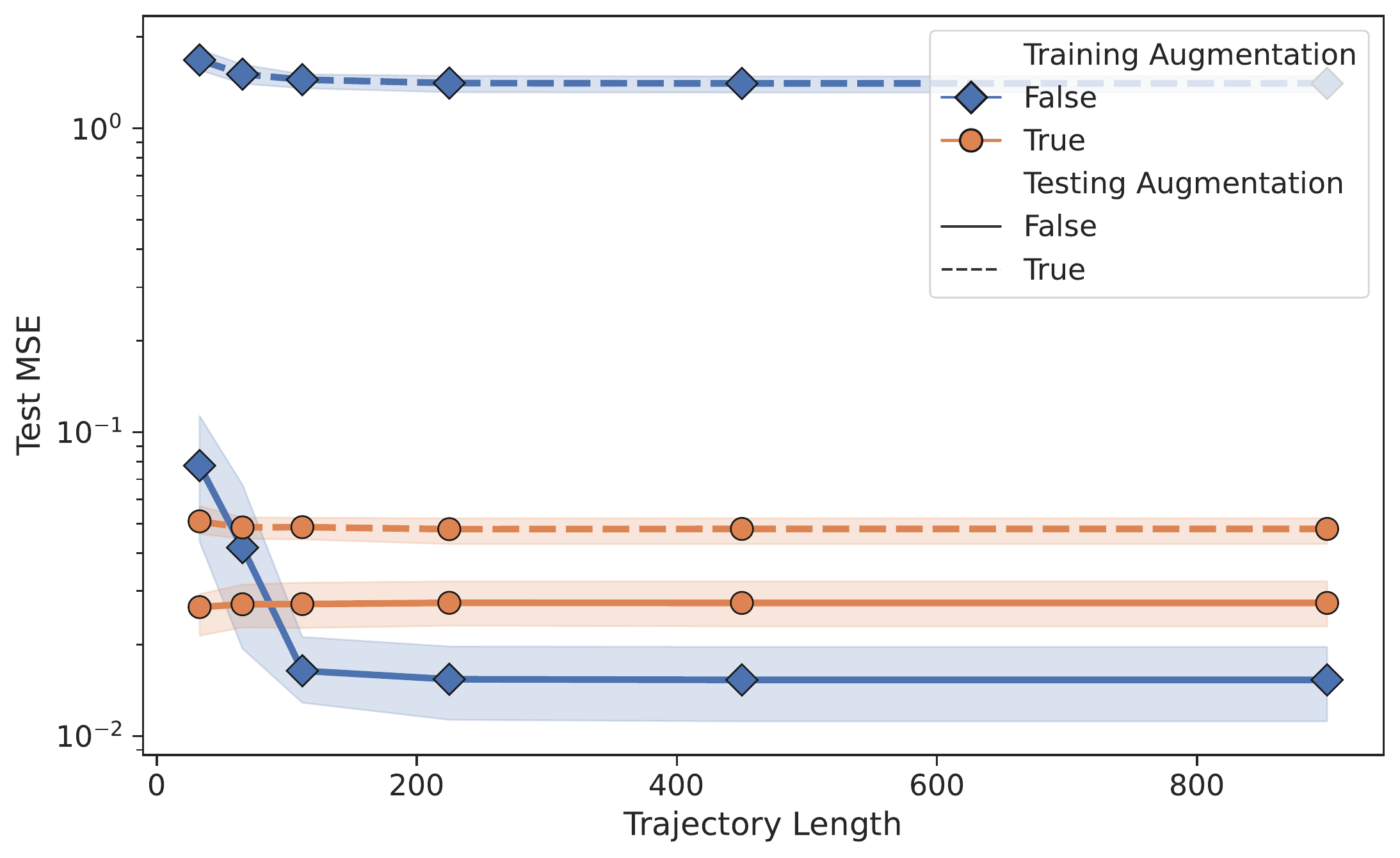}
    \end{center}
    \caption{Test error for a pair of augmented and non-augmented models, trained on a Hénon map dataset of 8172 input-output pairs for 25,000 optimizer steps. The non-augmented model was trained on trajectories with length $m=250$, whereas the augmented train data contained trajectories varying in length from $m=10$ to $m=250$. We then evaluated each model on test datasets which consisted of fixed-length trajectories, with and without test-time augmentation. For this experiment, the testing data augmentation consisted of taking a random subset of $n\in \{10, 11,\ldots, 225\}$ trajectories for each input-output pair. Error bands represent bootstrapped 95\% confidence intervals for models trained with three random seeds.}\label{fig:henon_generalization_study}
\end{figure}

\Cref{fig:henon_generalization_study} further explores the generalization gains observed from applying data augmentation to samples during model training. Both models evaluated in the figure appear to generalize across inputs with trajectories with lengths ranging from 200 to 1000 samples, achieving a roughly constant error across different trajectory lengths. The clearest difference between models in \Cref{fig:henon_generalization_study} is that the estimation error of the non-augmented model (diamond markers) grows up to 100 times larger on randomly augmented test inputs compared with the performance on inputs without this randomization (the dashed and solid lines respectively). In comparison, the augmented model displays a significantly smaller gap in performance on augmented versus non-augmented test data and, notably, demonstrated improved performance when tested on inputs without the random augmentation that the model was trained with. 

\begin{figure}[ht!]
    \begin{center}
        \includegraphics[width=1\columnwidth]{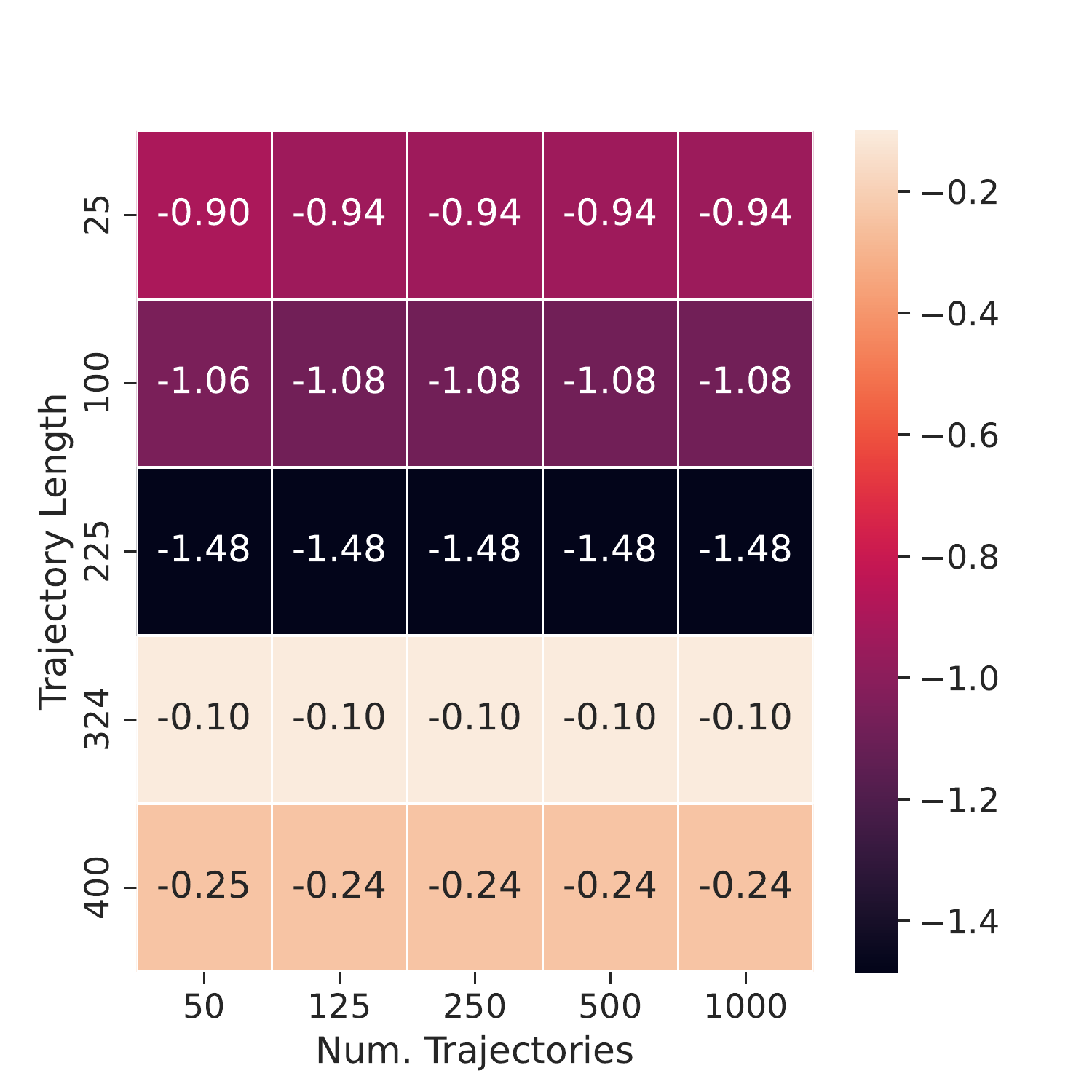}
    \end{center}
    \caption{A heatmap (darker is better) displaying the logarithm of the test error of the Hénon map augmented model model for 25,000 optimizer steps \Cref{fig:henon_generalization_study} (dashed orange curve), on test datasets in which the number of trajectories is varied, as well as the length of trajectories.
     }\label{fig:henon_two_axis_generalization}
\end{figure}

Finally, we examined the limits of the generalization abilities of augmented models. \Cref{fig:henon_two_axis_generalization} shows the results of an experiment in which we tested the estimation performance of such a model on test pairs with between 50 and 1000 length trajectories and between 25 and 400 trajectories per sample. As in \Cref{fig:henon_generalization_study}, we observed that the model generalized well to short and long trajectory lengths, including ones longer than the maximum length of 225 samples in the training set. The model's performance was more significantly affected by the number of trajectories per sample, with performance improving as the number of trajectories increased from 25 to 225 before a significant deterioration in performance on samples with 324 and 400 trajectories in each sample.

\subsection{The Swinging Atwood's Machine}\label{subsec:SAM}
The swinging Atwood's machine (SAM) \cite{Tufillaro1984} is a Hamiltonian system with nonlinear dynamics, consisting of an Atwood's machine in which one of the mechanical bobs is allowed to swing from its pulley along the two-dimensional plane containing the bobs and their pulleys. The configuration manifold of the system is two-dimensional, consisting of the distance between the swinging pendulum and its
pulley $r \in (0, \infty)$, and the angle formed between the swinging bob and vertical, $\phi \in (-\pi, \pi]$.

This system exhibits chaotic motion for nearly all values of the mass ratio $\mu \in (0, \infty)$ between the stationary and the swinging pendulum, taken as the parameter of the system \cite{Casasayas1990}. For our experiments, we set the gravitational acceleration $g$, and the system's mechanical energy $E$ to unity, as is common when studying this system. In our experiments, we used examples with $\mu \in [1.5,15]$. For ease of comparison with our Hénon map results, we used a rescaled loss function
\begin{equation}
    \ell\rbr{\hat{\mu}, \mu} = \dfrac{1}{2} \sigma_{\mu}^2 (\hat{\mu} - \mu)^2,
\end{equation}
where $\sigma_{\mu}=.\overline{740}$ was chosen so that $\sigma_{\mu} \cd (\mu - \mu_\text{mid})$ takes values in the interval $[-5,5]$, where $\mu_\text{mid}=8.25$ is the center of the parameter interval $[1.5,15]$.

\subsubsection{Choice of Poincaré Section}\label{subsubsec:SAM_poincare_section}
\begin{figure}[ht!]
    \begin{center}
        \includegraphics[width=.85\columnwidth]{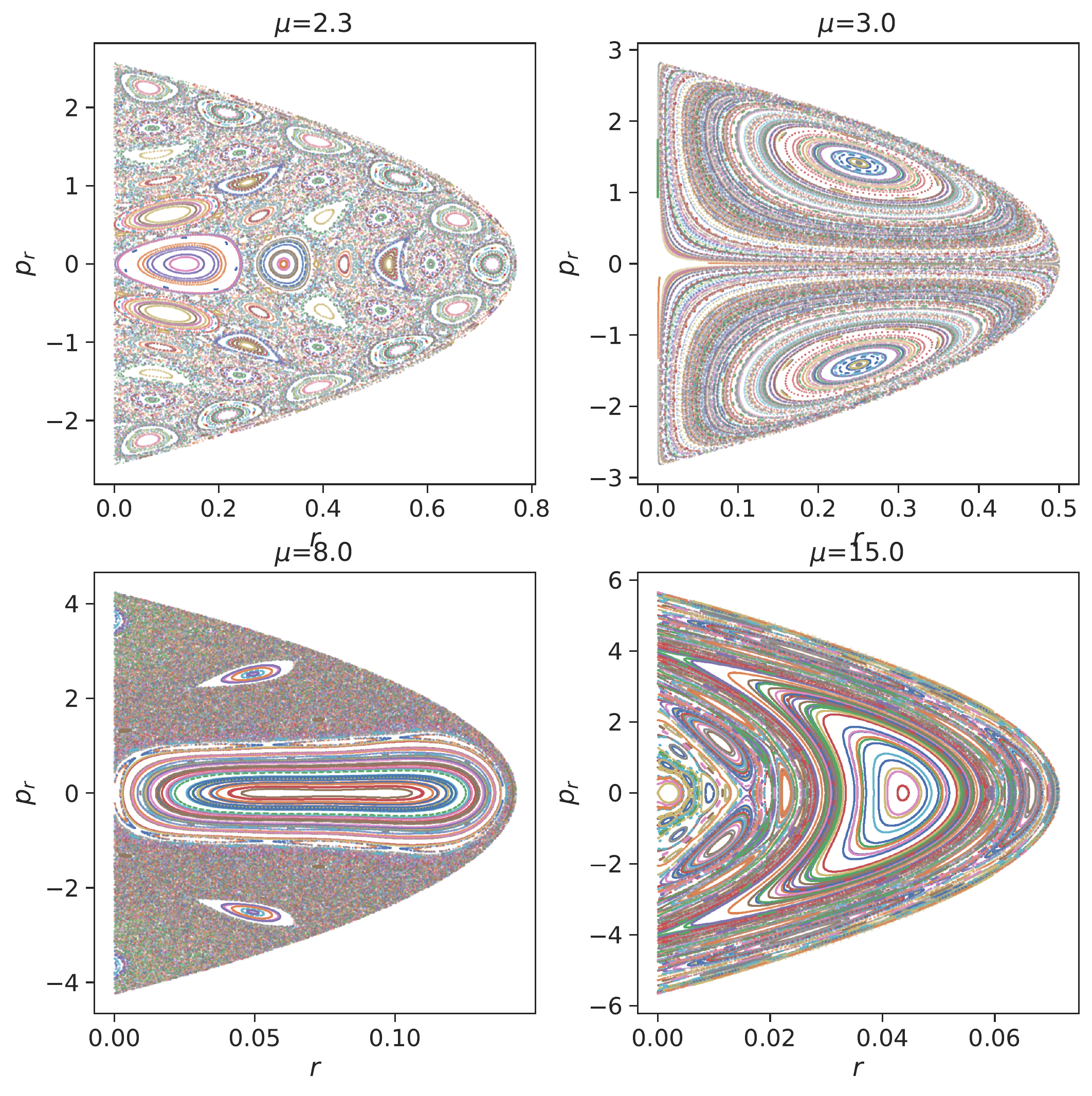}
        \caption{Return maps for the SAM system obtained using the Poincaré section described in \Cref{subsubsec:SAM_poincare_section}. The system was integrated forwards for 1000 units of time for 200 initial conditions, drawn randomly from energetically-allowed initial states.}\label{fig:SAM_portraits}
    \end{center}
\end{figure}

For our experiments we used a common Poincaré section for the SAM: the section where the state of the system $(r, \phi, \dot{r}, \dot{\phi})$ passes through $\phi=0$ with $\dot{\phi}>0$. \Cref{fig:SAM_portraits} shows return maps for this section corresponding to four different values of $\mu$, including $\mu=3$ for which the system is non-chaotic \cite{Tufillaro1984}.

Given that our choice of Poincaré section constrains $\phi=0$ and the Hamiltonian of the system is time-independent, the system's state at each section crossing is uniquely determined by the values of $r$ and its conjugate momentum $p_r$. We used two-dimensional return maps that only consider these values, e.g. the return maps in \Cref{fig:SAM_portraits}.

\subsubsection{Data Generation}\label{subsubsec:SAM_data_gen}
To create the dataset $\mathcal{D}$ for these experiments, we used 4,000 evenly spaced values for $\mu \in [1.5, 15]$. Because the SAM is a Hamiltonian system and we considered crossing points for which $\phi=0$, one can determine that all crossing points of the system lie within fixed values of $r, \phi$ for each value of the value of mass ratio parameter $\mu$ at a fixed mechanical energy of 1. The roughly triangular boundary is visible in the different portraits in \Cref{fig:SAM_portraits}. Using this fact, for each value of $\mu$ we used Heyoka to integrate 256 initial states randomly selected from the energetically allowed regions of the state space for 1,000 units of system time. For each initial state, we recorded the state of the system each time it crossed through the Poincaré section specified in \Cref{subsubsec:SAM_poincare_section}. This data generation method resulted in trajectories of varying lengths, depending on how many times each trajectory passed through the section within 1,000 units of time, and so in order to perform data augmentation we cropped each trajectory down from its original length to between 1 and 250 points.

\subsubsection{Results}\label{subsubsec:results}
\begin{figure}[ht!]
    \begin{center}
        \includegraphics[width=1\columnwidth]{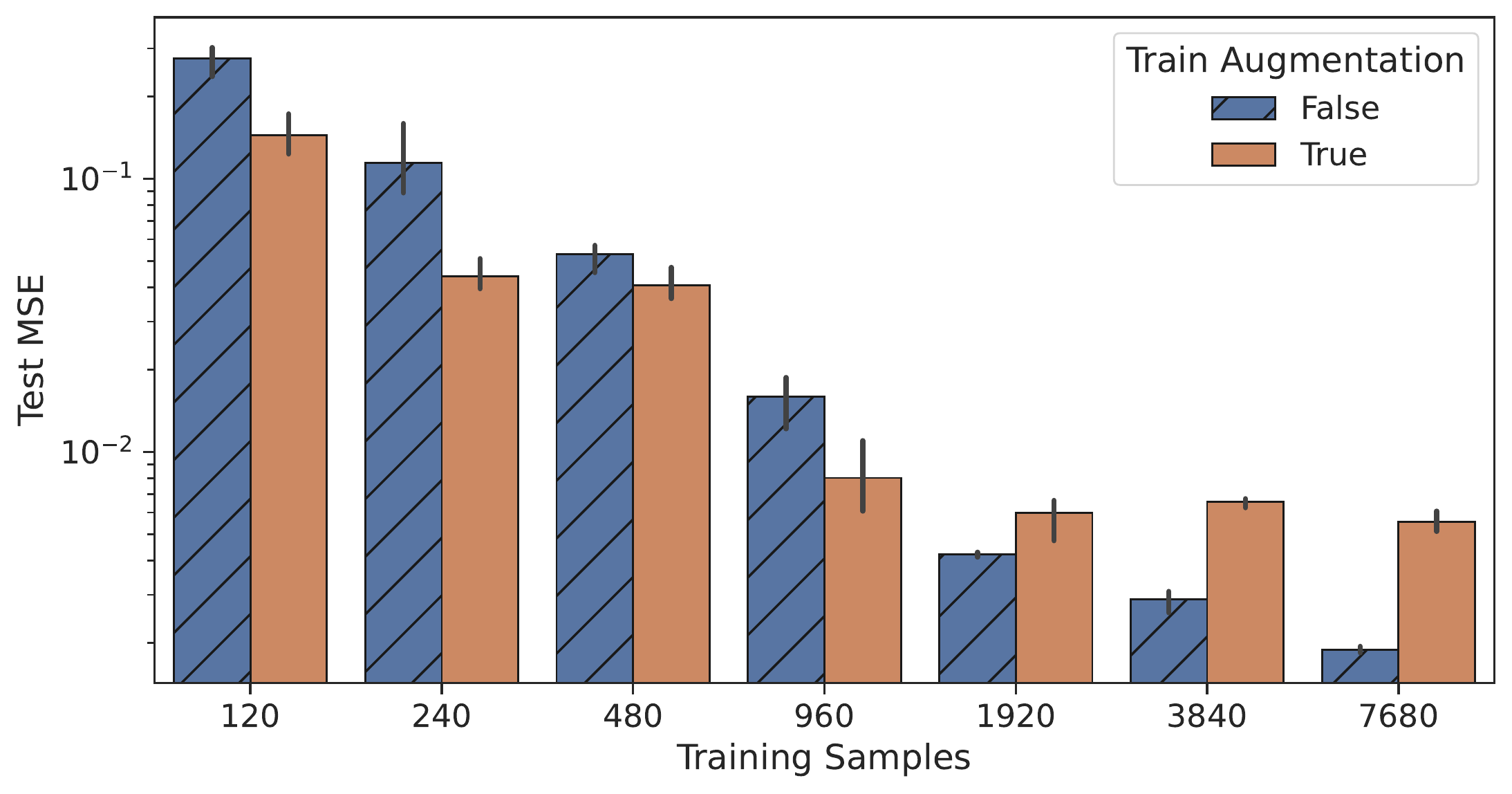}
    \end{center}
    \caption{Test mean squared error of augmented and non-augmented models on datasets of increasing numbers of sample pairs from the swinging Atwood's Machine system with the. Each model was trained for 25,000 optimization steps using the same configuration of the Adam optimizer (learning rate, weight decay, batch size). For each dataset size, the five model versions with the lowest validation error observed during training were selected. We then evaluated these models on a test set. We evaluated model errors on inputs that matched their respective training distributions. I.e., we only applied data augmentation to the test samples for augmented models. Error bars indicate bootstrapped $95\%$ confidence intervals from models trained with three different random seeds.}\label{fig:SAM_data_aug_test_barplot}
\end{figure}

\begin{figure}[ht!]
    \begin{center}
        \includegraphics[width=1\columnwidth]{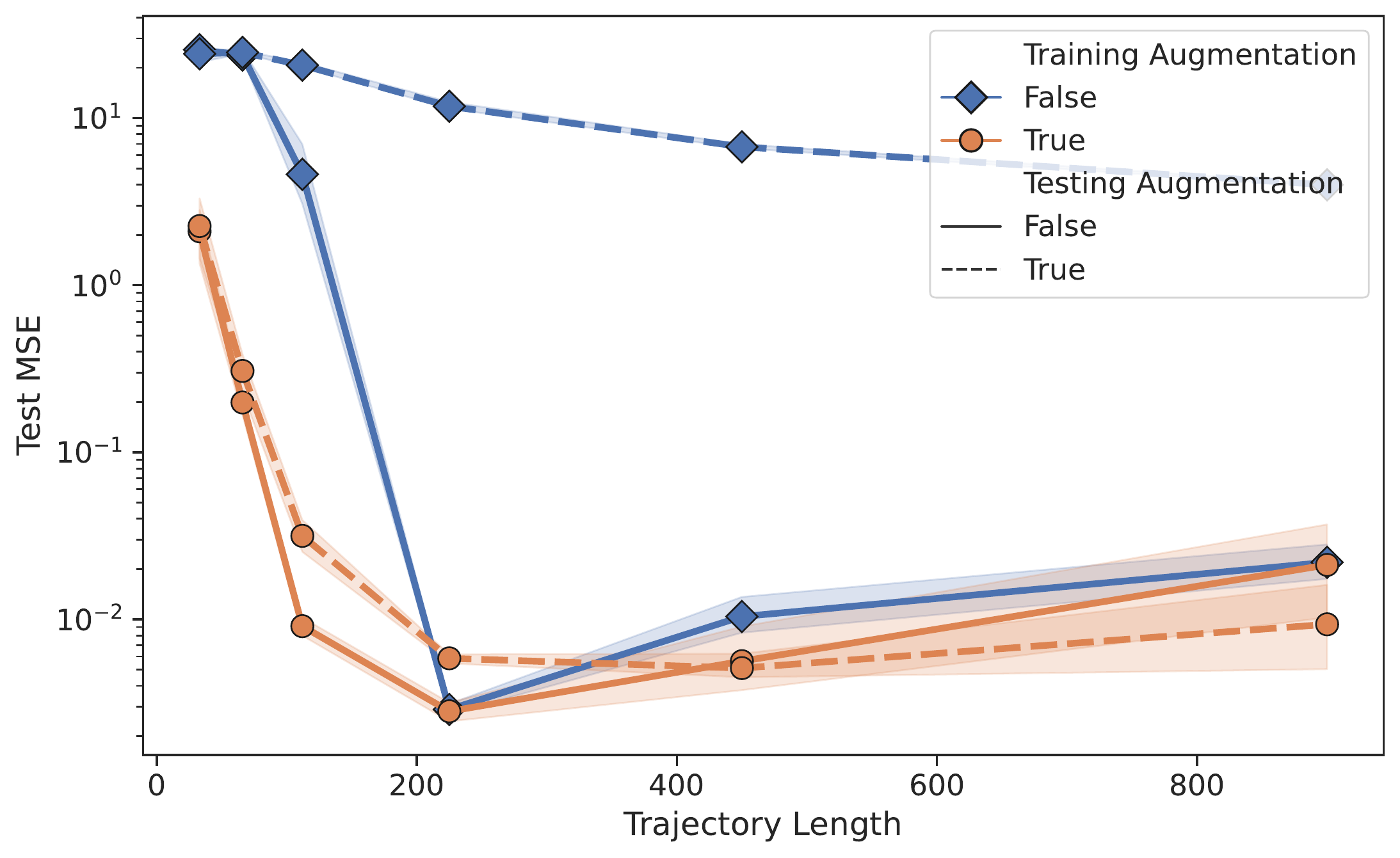}
    \end{center}
    \caption{
        Test error for a pair of augmented and non-augmented models, trained for 25,000 optimizer steps on the swinging Atwood's Machine consisting of 7680 input-output pairs. Both the non-augmented and augmented training trajectories varied in length from 1 to 250 points. We evaluated each model on test datasets which consisted of fixed-length trajectories, with and without test-time augmentation, which, for this experiment, consisted of taking a random subset of the trajectories for each input-output pair. the testing data augmentation consisted of taking a random subset of $n\in \{10, 11,\ldots, 225\}$ trajectories for each input-output pair. Error bands represent bootstrapped 95\% confidence intervals from models trained with three different random seeds.}\label{fig:SAM_generalization_study}
\end{figure}

The experiments in this section closely mirror those for the Hénon map system in \Cref{subsubsec:henon_results}. \Cref{fig:SAM_data_aug_test_barplot} summarizes the results of an experiment investigating the impacts of training dataset size and our data augmentation process for the swinging Atwood's Machine system. Similar to in \Cref{fig:henon_data_aug_barplot} we observe that the test error of our estimation models decreases in an approximately power-law fashion with the size of the training set, although we observe that the performance of the augmented model appears to plateau. Prior to this plateau, we see that the augmented models often perform better than non-augmented models.

\Cref{fig:SAM_generalization_study} shows the results of our experiment looking at the generalization benefit of applying data augmentation to samples during model training. In contrast with \Cref{fig:henon_generalization_study} we see that the performance of both models is worse on inputs with shorter trajectory lengths. Once again we observe that there is a smaller generalization gap for augmented model as evidenced by the vertical gap between the lines with circular markers versus those with diamond markers.


\section{Discussion}\label{sec:Discussion}
By framing PSI as a supervised machine learning problem, we circumvent the use of loss functions that compare simulated trajectories to observations. Instead, we directly learn estimation functions that minimize parameter estimation errors, which is a natural objective for parametric system identification. In this setting, the ability to obtain effective estimation functions from potentially limited datasets becomes a central concern. In our experiments, we observed an approximately power-law reduction in estimation errors by increasing the number of training samples used. This is a common phenomenon in statistical estimation problems \cite{Wainwright2019} and implies that the availability of training data is an important factor in the success of our method. 

With that said, we observed that for small sample sizes, there were significant improvements in estimation accuracy when using random data augmentation during model training. While the reason for the empirical success of data augmentation is an open area of research (see e.g. \cite{Shen2022}), one explanation for our observations is that data augmentation effectively increases the number and variety of training samples used to optimize the estimation model's parameters, resulting in a model which generalizes better to new samples. On the other hand, we also observed faster improvement in model performance with the size of the training set for non-augmented models, with the test MSE of these models overtaking their augmented counterparts on both dynamical systems in this paper. On this last point, while the test error of non-augmented models on larger training datasets surpassed that of non-augmented models, we would argue that the augmented models still have more favourable characteristics for a practical model, as we discuss next.

\Cref{fig:henon_generalization_study,fig:SAM_generalization_study} provide a more complete characterization of the performance of augmented and non-augmented models trained on the largest datasets shown in \Cref{fig:henon_data_aug_barplot,fig:SAM_data_aug_test_barplot} -- 8172 and 7680 samples, respectively. In these line plots, we observe that the augmented models have improved performance across input distributions of varying quality, especially on samples with shorter trajectories. This can be seen in both the smaller gap between test errors on augmented and non-augmented test samples and in the observation that the best-performing models on test inputs with the shortest trajectories were those trained with random data augmentation (left-hand sides of \Cref{fig:henon_generalization_study,fig:SAM_generalization_study}). These results indicate that the use of these augmentation methods is currently a useful tool for building practical data-driven estimation models.

All of our experiments were performed with $\alpha=.7$ and 128 by 128-pixel images. These `hyperparameters', as they are referred to in the machine learning community, can all in principle be optimized on validation data, as we did with the batch size and parameters of the optimization routine, at the cost of increased computation time.


A consideration for the method we propose is that we rely on changes to system parameter values translating into changes to the resulting return maps. If large changes to parameter values result in only small changes to return maps then we would expect our method to have a large estimation error. This behavior should also occur with all previous methods that rely on loss functions that compare observed and simulated trajectories in state space. In addition, in most scenarios, this effect is likely benign, given that in such a scenario a large estimation error may be acceptable since trajectory predictions and system analysis relying on estimated parameters would remain reasonably accurate due to the low sensitivity of the dynamical system to changes in its parameter values. 

Continuous-time dynamical systems pose an additional challenge, in that if the chosen Poincaré section fails to adequately capture key features of trajectories, then by unfortunate coincidence two different parameter values may result in different dynamics but share the same Poincaré map. This issue is separate from a possible issue with the collected data in which all parameter values for a system have distinct Poincaré maps for a given Poincaré section, but there are not enough data points collected for each parameter value to disambiguate between parameter values in some cases. Instead, we refer to a scenario where for distinct parameter values $\theta$ and $\theta'$, the dynamics of the system result in the same Poincaré map but different trajectories away from the Poincaré section, in which case no amount of data collected on such a Poincaré section can distinguish between systems parameterized by $\theta$ versus $\theta'$.

This consideration means that care should be taken to ensure that the choice of Poincaré section used to study a dynamical system captures the dynamics of interest. Our method straightforwardly accommodates using two or more return maps for each input pair, for example, return maps corresponding to different Poincaré maps for the same parameter value, by stacking each return map as an image channel before input to the ResNet parameter estimation model. This method effectively provides the estimation model with different `cross-sections' of trajectory dynamics to use for parameter estimates and may lead to better estimates if one of the Poincaré sections is not informative for some parameter values.



\section{Conclusion}\label{sec:Conclusion}
In this work, we have introduced a novel solution method for parametric system identification (PSI), framing the problem of mapping sample trajectories to system parameters as a supervised machine learning problem. Combining this idea with recent approaches to PSI, which use state-space representations of trajectories, we show that our approach is effective for parameter identification on chaotic dynamical systems. Since we use return maps as input to a supervised machine learning model, our method generalizes to continuous time dynamical systems using Poincaré maps. Although training the estimation models is a compute-intensive process, the resulting models provide fast, inexpensive, and accurate parameter estimates that can be used to process large collections of data or as subcomponents of a more complicated program. While we focused on systems with two-dimensional in this work, our method extends to more dimensions simply by replacing the two-dimensional pixelized transformation and subsequent convolution operations in the ResNet with their higher dimensional analogs. 

This work opens the door to future exploration of machine learning for PSI. In our current method, we effectively reduce the problem of PSI to a computer vision task, and so there are likely performance improvements, both in terms of estimation accuracy and in terms of reducing the number of samples required to reach high accuracy by more carefully optimizing the estimation architecture used. One possible avenue is to use the wide availability of large, pre-trained deep learning models trained on image tasks with more layers than our ResNet18 model and only optimize the final layers of the network for the parameter estimation task. This `transfer learning' \cite{Weiss2016} method can potentially reduce the number of samples required to train a new estimation model.

A second, more open direction is to investigate more sophisticated methods for structuring return maps as input to a deep learning model. Our approach of flattening these maps into single-channel images introduces several extraneous parameters. In addition, pixelating return maps potentially removes some of the finer spatial structure present in the return map when using coarser pixelation schemes. An alternative approach could be to explicitly represent return maps as unstructured collections of points and use a graph neural network \cite{Scarselli2009} as the parameter estimation model.


Deep learning and other data-driven modeling and optimization methods such as reinforcement learning are becoming increasingly useful tools in the physical sciences. While these approaches can be powerful, there are often significant challenges associated with developing useful applications in different problem domains. In this work, we take a step in the right direction, and we invite the research community to explore the possibilities of using data-driven methods and return maps to analyze complex dynamical systems.



\backmatter


\bmhead{Acknowledgments}

\section*{Declarations}\label{sec:Declarations}
\subsection*{Funding}
The authors declare that no funds, grants, or other support were received during the preparation of this manuscript.
\subsection*{Conflict of interest/Competing interests}
The authors have no relevant financial or non-financial interests to disclose.
\subsection*{Ethics approval}
Not applicable.
\subsection*{Consent to participate}
Not applicable.
\subsection*{Consent for publication}
Not applicable.
\subsection*{Availability of data and materials}
See the subsection `Code Availability'.
\subsection*{Code availability}\label{sec:code_avail}
We have included the code to replicate our experiments \href{https://github.com/connorsteph/parameter_regression_from_return_maps_paper_code}{here}. The code repository contains the configuration files used to generate the SAM dataset files, as well as the configuration files we used to train the models on the SAM and Hénon map systems shown in this paper.
\subsection*{Authors' contributions} 
Emmanuel Blazquez provided project supervision and editorial support in writing the manuscript. Connor James Stephens developed the code used to run the experiments in this paper and wrote the manuscript.

\noindent

\bigskip





\begin{appendices}




\end{appendices}


\bibliography{bibliography_sync}


\end{document}